\documentclass[a4paper,twoside,notoc,11pt]{JHEP3}
\pdfoutput=1
\usepackage{epsfig,multicol,slashed}
\usepackage{delarray,amsmath,bbm}

\newcommand\fverb{\setbox\pippobox=\hbox\bgroup\verb}
\newcommand\fverbdo{\egroup\medskip\noindent%
                              \fbox{\unhbox\pippobox}\ }
\newcommand\fverbit{\egroup\item[\fbox{\unhbox\pippobox}]}
\newbox\pippobox
\newcommand{\nn}{\nonumber}
\newcommand{\beq} {\begin{equation}}
\newcommand{\eeq} {\end{equation}}
\newcommand{\beqa} {\begin{eqnarray}}
\newcommand{\eeqa} {\end{eqnarray}}

\newcommand{\ie}{{\it i.e.}}

\newcommand{\cf}{{\it cf.\ }}

\newcommand{\lqcd}{\Lambda_{QCD}}

\newcommand{\ieps}{i\varepsilon}

\newcommand{\veps}{\varepsilon}

\newcommand{\bk}{{\boldsymbol{k}}}

\newcommand{\bp}{{\boldsymbol{p}}}

\newcommand{\bqt}{{\boldsymbol{q}_\perp}}
\newcommand{\order}[1]{${\cal O}\left(#1 \right)$}
\newcommand{\morder}[1]{{\cal O}\left(#1 \right)}
\newcommand{\eq}[1]{(\ref{#1})}
\newcommand{\fig}[1]{Fig.~\ref{#1}}

\newcommand{\inv}[1]{\frac{1}{#1}}
\newcommand{\ket}[1]{\vert{#1}\rangle}
\newcommand{\bra}[1]{\langle{#1}\vert}

\newcommand{\acom}[2]{\left\{{#1},{#2}\right\}}

\newcommand{\tim}[1]{{\rm T}{\left\{ {#1} \right\}}}

\newcommand{\la}{\lambda}
\newcommand{\bs}[1]{\boldsymbol{#1}}
\newcommand{\pit}{(2\pi)^3}
\newcommand{\pif}{(2\pi)^4}

\newcommand{\kvec}{{\bs{k}}_\perp}
\newcommand{\kveco}{{\bs{k}}_{1\perp}}
\newcommand{\kvect}{{\bs{k}}_{2\perp}}

\newcommand{\lveco}{{\bs{\ell}_{1\perp}}}
\newcommand{\lvect}{\bs{\ell}_{2\perp}}
\newcommand{\notvec}{\bs{0}_\perp}

\newcommand{\half}{\frac{1}{2}}

\newcommand{\halft}{{\textstyle \frac{1}{2}}}

\newcommand{\gsim}{\gtrsim}

\newcommand{\aslash}[1]{ \rlap{/}{#1} }

\title{\center{Factorization at fixed $Q^2(1-x)$}}
\author{Paul Hoyer$^a$, Matti J\"arvinen$^b$ and Samu Kurki$^a$\\
              $^a$Department of Physics and Helsinki Institute of
              Physics\\
              \ POB 64, FIN-00014 University of Helsinki, Finland \\
              $^b$High Energy Physics Center, University of Southern Denmark,
			  Campusvej 55, DK-5230 Odense M, Denmark}
\preprint{HIP-2008-26/TH}

\abstract{We consider QCD factorization between hard and soft subprocesses in inclusive reactions where the momentum fraction $x$ of one parton approaches unity as the hard scale $Q^2 \to \infty$, such that $Q^2(1-x)$ is fixed. In this ``BB limit'' the entire (multi-parton) Fock state containing the high $x$ parton is coherent with the hard subprocess. The soft contribution is given by a forward multiparton matrix element. The BB limit corresponds to a fixed (large or small) missing mass and is thus closely connected to exclusive production. We analyze the Drell-Yan process $h+ N \to \gamma^* +X$ in detail, explaining why the virtual photon is longitudinally polarized for $h=\pi$ and transversely polarized for $h=p$. The BB limit may be relevant also for other phenomena observed at high $x$, such as the large single spin asymmetries of $pp\to \Lambda^\uparrow X$ and in $pp^\uparrow \to \pi X$.
}


\begin{document}

\section{The BB limit}

Data on hard inclusive processes has been successfully analyzed assuming QCD factorization between a hard subprocess and universal soft matrix elements (parton distributions). Formally one considers the leading contributions in the Bj limit where a hard scale $Q^2 \to \infty$ while the momentum fractions $x_i$ carried by the active partons (one in each hadron) are held fixed. The higher twist corrections are power suppressed in the hard scale but generically increase as $x_i \to 1$. Thus the effective expansion parameter is $1/[Q^2(1-x_i)]$ (see \cite{Blumlein:2008kz} for a recent  phenomenological analysis of $eN \to eX$ (DIS)). 

Data at high $x$ has features which differ qualitatively from the leading twist contribution. A striking example is the polarization of the virtual photon in $\pi N \to \mu^+ \mu^- X$ which changes from transverse to longitudinal when the photon carries a momentum fraction $x_F \gsim 0.6$ of the pion beam \cite{Anderson:1979xx}. This requires that at least one of the annihilating quarks in the subprocess $q\bar q \to \gamma^*(Q^2)$ goes off-shell by an amount commensurate with the photon virtuality  $Q^2$. As pointed out by Berger and Brodsky \cite{Berger:1979du} the longitudinally polarized photon is coherent with both valence quarks in the pion. Hence the dominant contribution is of higher twist even though the subprocess is hard. An analogous change of $J/\psi$ polarization at high $x_F$ was observed in $\pi N \to J/\psi+ X$ \cite{Biino:1987qu}. The inapplicability of the twist expansion in QCD at high $x$ was discussed in \cite{Brodsky:1991dj}.

In this paper we consider QCD factorization in a limit where $x \to 1$ as $Q^2 \to \infty$. Here $x$ can denote either the momentum fraction $x_i$ of 
a fast quark $i$ or equivalently the $x_F$ of a final state particle.
The Drell-Yan data suggests that the helicity of the pion is transferred to the virtual photon at high $x$, implying that the photon is coherent with an entire Fock state of the pion. The life-time of a Fock state is inversely proportional to $\Delta E$, the energy difference between the pion and the Fock state. At high pion momentum $p$, 
\beq\label{endiff}
2p\Delta E \simeq m_\pi^2-\sum_i\frac{p_{i\perp}^2+m_i^2}{x_i}
\eeq
The $x \simeq 1$ quark which annihilates into the virtual photon contributes only little to the energy difference \eq{endiff}. The life-time of the Fock state is determined by the stopped partons with $x_i \propto 1-x$. If  $1-x \sim \lqcd^2/Q^2$ we get $2p\Delta E \sim Q^2$ (we take $p_{i\perp} \sim \lqcd$, a generic soft QCD scale). Such Fock states have life-times similar to that of the virtual photon, ensuring their coherence with the hard process. This motivates us to consider the
\beq\label{bblimit}
{\rm BB\ limit:}\ \ \ Q^2 \to \infty \ \ \ {\rm at\ fixed}\ \ Q^2(1-x)
\eeq
which recognizes \cite{Hoyer:2006hu} the early observations of Berger and Brodsky \cite{Berger:1979du}. 

The understanding of QCD factorization in the BB limit is facilitated by an analogy with ordinary DIS $(ep \to eX)$. In the rest frame of the target DIS may be viewed \cite{AlignedJet} as proceeding through a splitting of the virtual photon into a quark pair, $\gamma^*(Q^2) \to q\bar q$. The quark carries nearly all of the photon energy $\nu$ and forms the current jet, whereas the antiquark carries a finite momentum in the target rest frame, even as $\nu = Q^2/2m_px_B \to \infty$. The $q\bar q$ Fock state of the photon is analogous to the pion Fock state we just discussed: A fast quark with momentum fraction $x_q \simeq 1$ and an antiquark with $x_{\bar q} \sim \lqcd^2/Q^2$. The fact that the momentum $x_{\bar q}\nu$ of the antiquark remains finite in the Bj limit allows the alternative (standard) interpretation of the antiquark as a quark constituent of the proton. The DIS cross section can then be equivalently ascribed to the scattering of the $q\bar q$ Fock state on the target
or to the probability of finding a quark in the target wave function. The latter interpretation corresponds to the parton distributions obtained in QCD factorization. 

In the following we shall analyze factorization of the Drell-Yan process in the BB limit, interpreting stopped partons in Fock states of the beam as (anti-)parton constituents of the target. We do not consider rescattering effects, \ie, gauge links between the quark fields \cite{Brodsky:2002ue}.

\section{Kinematics of $\pi^+ N \to \gamma^* + X$ at fixed $(1-x_F)Q^2$}

Our notation for the $\pi^+ N \to \gamma^* + X$ kinematics is indicated in the Feynman diagram of Fig.~\ref{fig1}. The pion, nucleon and photon momenta are in Light-Front (LF) notation ($k=(k^+,k^-,\kvec)$ with $k^\pm = k^0\pm k^3$)
%
\EPSFIGURE[hr]{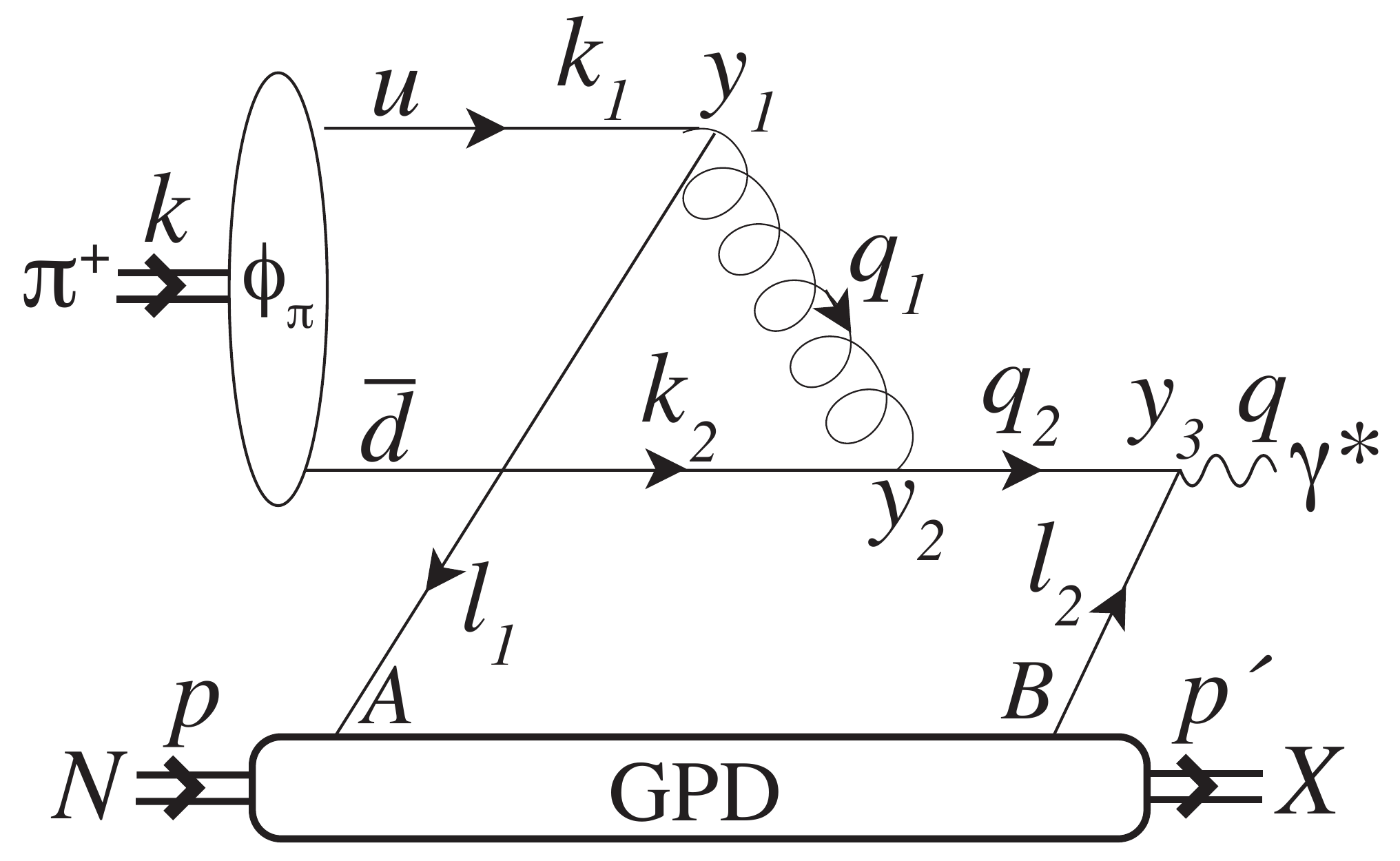,width=0.5\columnwidth}{Feynman diagram contributing to the $T_d(\pi N\to \gamma^* X)$ amplitude. 
The arrows indicate momentum directions; the $y_i$ are space-time positions of the vertices and $A,B$ denote color indices. The pion contributes via its $u\bar d$ distribution amplitude $\phi_\pi$ and the GPD denotes a transition ($N \to X$) Generalized Parton Distribution.
\label{fig1}}
\vspace{-6mm}
\beqa\label{momdef}
{\rm Pion:\ \ }k&=&(0,k^-,\bs{0}_\perp)\nn\\
{\rm Nucleon:\ \ }p&=&(p^+,m_N^2/p^+,\bs{0}_\perp)\nn\\ 
{\rm Photon:\ \ }q &=& (Q^2/q^-,q^-,\bs{q}_\perp)\ \ 
\eeqa
We neglect the mass of the pion as we take its momentum $k^- \to \infty$. We work in the target rest frame, $p^+ = p^- = m_N$, thus $s=(k+p)^2 \simeq m_N k^- \to \infty$. 
In the BB limit $x_F \equiv q^-/k^- \to 1$ at fixed $\bs{q}_\perp$, keeping also
\beqa\label{bblimit2}
x_B &\equiv& \frac{q^+}{p^+} = \frac{Q^2}{s}\ \ {\rm fixed}\nn\\ \\
k^- - q^- &\equiv& x_M p^- = \frac{Q^2(1-x_F)}{q^+} \ \ {\rm fixed}\nn 
\eeqa

\noindent Since the four-momentum transfer $k-q$ to the target system is fixed the invariant mass $M_X$ of the hadronic state $X$ is finite in the BB limit,
\beq\label{xmass}
M_X^2 = (k+p-q)^2 \simeq (1-x_B)[s(1-x_F)+m_N^2]-\bs{q}_\perp^2
\eeq
and may be small or large depending on the value of 
$s(1-x_F) = Q^2 (1-x_F)/x_B$.
For $X=N$ we have the exclusive Drell-Yan process $\pi N \to \gamma^* N$ \cite{Berger:2001zn} as well as the time-reversed version of $\gamma^* N \to \pi^+ N$ (Deeply Exclusive Meson Production) which are well-known to be described by Generalized Parton Distributions \cite{Diehl:2003ny} as indicated in Fig.~\ref{fig1}. Factorization  \cite{Collins:1996fb} applies equally to the transition GPD's ($X \neq N$) \cite{Frankfurt:1999fp,Pire:2005ax} since $M_X \ll Q$. In section 3 we merely sketch the derivation of the factorized amplitude. 

In section 4 we consider the factorization of the inclusive Drell-Yan cross section $\sigma(\pi^+N \to \mu\mu X)$ through a completeness sum over the states $X$. As in the standard leading twist limit of the Drell-Yan process, this includes integrating over the (small) transverse momentum $\bs{q}_\perp$ of the virtual photon, which arises from the intrinsic momentum of the partons in the GPD. High transverse momenta are suppressed at fixed $M_X$ since the GPD is expected to fall off $\propto 1/\bs{q}_\perp^4$, similarly to nucleon transition form factors. The integration over $\bs{q}_\perp$ can be done at fixed $M_X$ by varying $(k-q)^-$ without affecting the hard subprocess
at leading order.

We anticipate that the pion will contribute through its distribution amplitude $\phi_\pi(z)$ and parametrize the momenta of its valence quarks as
\beqa\label{kmom}
k_1 &=& \left(0,zk^-,\kvec \right)\nn\\
k_2 &=& \left(0,(1-z)k^-,-\kvec \right)
\eeqa
We take the transverse momenta to be limited, $\bs{k}_\perp^2 \ll Q^2$. Thus we neglect the perturbative tail of the wave function which arises from gluon exchange and gives rise to the logarithmic $Q^2$ evolution of the distribution amplitude 
\cite{Efremov:1979qk}. $k_1^+$ and $k_2^+$ vanish in the high energy limit and do not contribute at leading order.

In the BB limit \eq{bblimit} one of the quarks in the pion (the $u$-quark in Fig.~\ref{fig1}) transfers nearly all its momentum to the other quark, such that the photon carries nearly all of the pion momentum. The stopped $u$-quark is left with a finite momentum in the target rest frame and should be connected to the target wave function as discussed above for DIS. Consequently, as $k^- \to \infty$,
\beqa\label{xdef} 
x &=& l_1^+/p^+\ \ {\rm fixed}\nn\\ 
x_B+x &=& l_2^+/p^+\ \ {\rm fixed}
\eeqa
A large longitudinal momentum is transferred through the gluon $(q_1)$ and quark $(q_2)$ propagators whose virtualities are of \order{Q^2},
\beq\label{ellvirt}
\left.
\begin{minipage}{3cm}\vspace{-.7cm}
\beqa
q_1^2 &\simeq& -zk^-l_1^+\nn\\
q_2^2 &\simeq& -k^-l_1^+\nn
\eeqa
\end{minipage}
\right\} \propto Q^2 = x_B s 
\eeq
This implies that the hard interactions occur at nearly the same LF times $(y_1^+,\ y_2^+,\ y_3^+=\morder{1/k^-})$
and vanishing transverse separations $(y_{1\perp},\ y_{2\perp},\ y_{3\perp}=\morder{1/Q})$. 

\EPSFIGURE[hr]{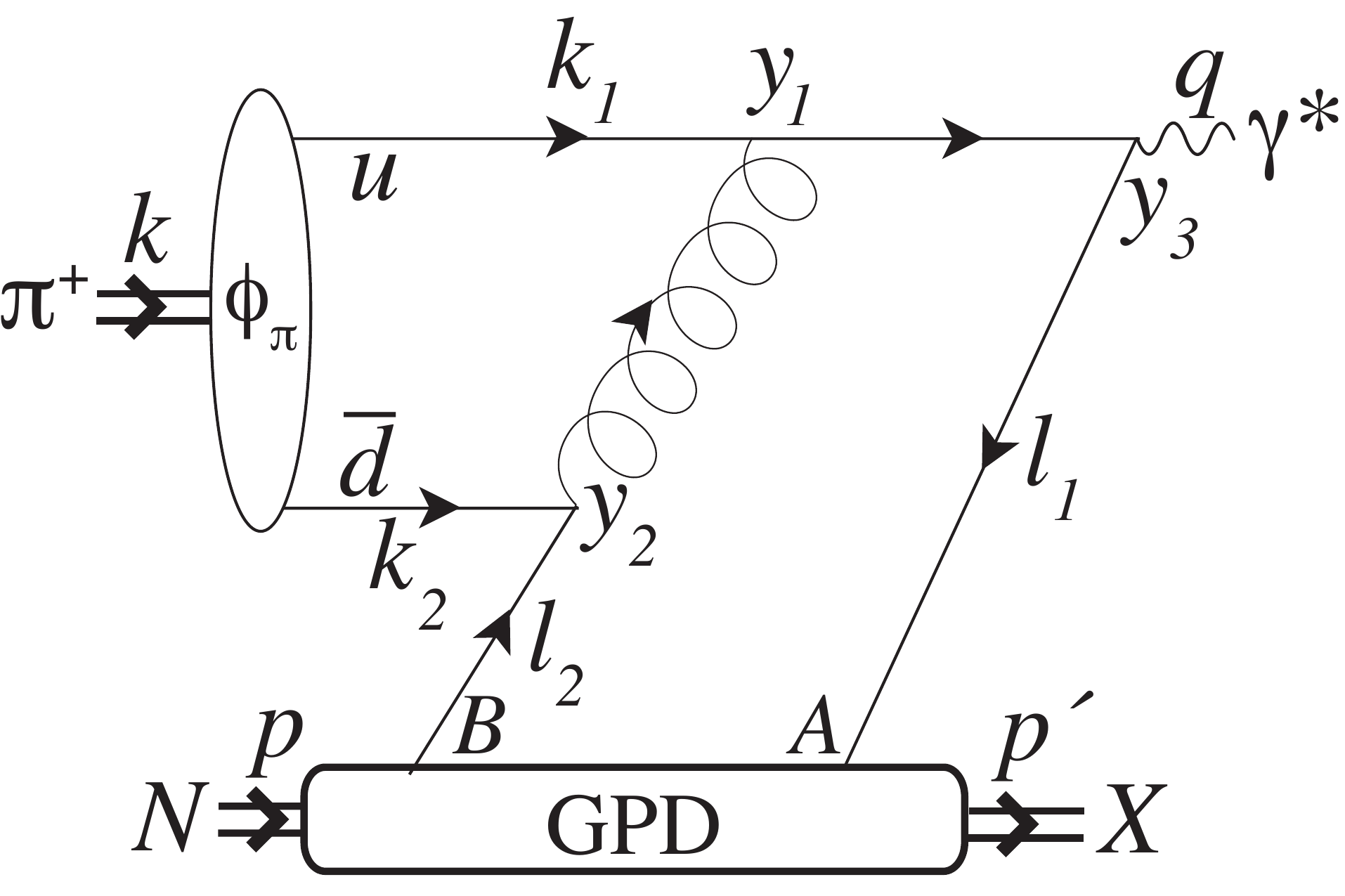,width=0.5\columnwidth}{Feynman diagram contributing to $T_u(\pi^+ N\to \gamma^* X)$. 
Notations as in Fig.~1.
\label{fig2}}

Since $q_1^2$ and $q_2^2$ are independent of $l_i^-$ and $l_{i\perp}$ the soft target matrix element will be integrated over these components (with the kinematic constraint 
$l_1-l_2=k-q$). In evaluating the hard, perturbative subprocess it suffices to take
\beqa \label{lmom}
l_1 &=& (xp^+,0^-,\bs{0}_\perp)\nn\\
l_2 &=& ((x+x_B)p^+,0^-,\bs{0}_\perp)
\eeqa
and analogously only the minus components in \eq{kmom} contribute. Thus both quarks effectively reverse their direction of motion along the $z$-axis.

\break

\section{Drell-Yan amplitude in the BB limit}

The general expression for the scattering amplitude is
\beqa\label{scatamp}
&&iT(\pi^+ N \to \gamma_\lambda^* X) (2\pi)^4 \delta^4(k+p-q-p') \nn\\ &&=\bra{\gamma_\lambda^* X}\tim{\exp[-i\int dt H_I(t)]}\ket{\pi N}\nn\\
\eeqa
The expression for a diagram such as in \fig{fig1} is obtained by expanding the hard 
vertices of the interaction Hamiltonian $H_I$, 
connecting them by perturbative propagators and retaining the leading contribution

\noindent in the BB limit \eq{bblimit2}. Equivalently, we may simply note that due to the large momenta of \order{k^-} flowing from $y_1 \to y_2 \to y_3$ the corresponding LF time differences $y_2^+-y_1^+$ and $y_3^+-y_2^+$ are of \order{1/k^-}. 
Similarly, the squared transverse separations vanish as the inverse virtualities \eq{ellvirt}: $|\bs{y}_{2\perp} - \bs{y}_{1\perp}|$ and $|\bs{y}_{3\perp}-\bs{y}_{2\perp}|$ are of \order{1/Q}. 
The diagram may then be evaluated using standard Feynman rules with the prescription \eq{pirule} (see Appendix~A) for the pion. The quarks $l_1$ and $l_2$ are treated as external particles and their free wave functions are replaced according to
\beqa \label{effrules}
\bar u(l_1) &\to& \int\frac{dl_1^+}{2\pi}\frac{dy_1^-}{2}\bra{X(p')}\bar
\psi_u(y_1)\exp(-\halft iy_1^-l_1^+)\nn\\
u(l_2) &\to& \psi_d(0)\ket{N(p)}
\eeqa
Our conventions for the wave functions and operators are explained in detail in Appendix~A.
Adding the diagram where the gluon vertex $y_2$ in \fig{fig1} is on the $l_2$ line
we find for the amplitude where a photon of helicity $\lambda=0$ is emitted from the $d$-quark,
\beqa\label{amp4}
T_d(\pi^+ N\to\gamma^*_L X) = e_d\frac{-ieg^2\,C_F}{Q\sqrt{2N_c}}\hspace{8cm} \\
\times
\int \frac{dz\,dl_1^+\,\phi_\pi(z)}{z\, 2\pi (l_1^+ -\ieps)} \int dy_1^- \, e^{-iy_1^-l_1^+/2} \bra{X(p')}\bar\psi_u(y_1)\gamma^+
\gamma_5\,\psi_d(0) \ket{N(p)}_{y_1^+ = y_{1\perp}=0}\nn
\eeqa
where $\phi_\pi(z)$ is the pion distribution amplitude.
The amplitude for transversely polarized photons is suppressed by a factor $1/Q$. We give an intuitive explanation of this in section~5. 

The contribution from the diagrams where the photon is emitted from the $u$-quark 
as in \fig{fig2} is obtained in a similar way. Defining
\beq\label{Cdef}
C(x_B,x) \equiv  \int_0^1 dz\,\phi_\pi(z)  \left(\frac{e_u}{1-z}\inv{x_B+x +\ieps}+\frac{e_d}{z}\inv{x -\ieps}\right)
\eeq
the Drell-Yan amplitude for longitudinal photons is
\beqa\label{amp9}
T(\pi^+ N\to\gamma^*_L X) = T_u+T_d &=& \frac{-ieg^2\,C_F}{2\pi Q\sqrt{2N_c}}\int dx\, C(x_B,x) \nn \\ \\
&\times& \int dy_1^- \, e^{-iy_1^-l_1^+/2} \bra{X(p')}\bar\psi_u(y_1)\gamma^+\gamma_5\,\psi_d(0) \ket{N(p)}_{y_1^+ = y_{1\perp}=0}\nn
\eeqa
where $x$ and $x_B$ are defined in \eq{xdef}. For $X=N$ we may verify that the amplitude indeed corresponds to the usual expression for deeply exclusive pion production \cite{Collins:1996fb}.
For general states $X$ the matrix element is a ``transition'' GPD \cite{Frankfurt:1999fp,Pire:2005ax}.

The full amplitude including the muon vertex is
\beqa\label{ampmu}
T(\pi N\to \mu^+\mu^- X) &=& T(\pi N\to \gamma_L^* X) \frac{i}{Q^2} \bar u(q_1,s)(-ie)\slashed{\veps}_0(q) v(q_2,-s') \nn\\
&=&  T(\pi N\to \gamma_L^* X)\frac{ie\sin\theta}{Q}\delta_{ss'}
\eeqa
where $\theta$ is the polar angle of the muon momentum in the muon pair rest frame and $s,s'$ are the muon helicities.

\section{The Inclusive Drell-Yan cross section}

The inclusive cross section is obtained by summing over all final states $X$,
\beq\label{sig1}
\sigma(\pi^+ N\to\gamma^*_L X) = \inv{2s}\sum_X \int \frac{dq^- d^2\bqt}{\pit 2q^-}\, |T(\pi^+ N\to\gamma^*_L X)|^2 \pif \delta^4(k+p-q-p')
\eeq
where $\sum_X$ is defined by the completeness relation,
\beq\label{comp}
\sum_X \ket{X}\bra{X} \equiv \sum_{n=0}^\infty\int\prod_{i=1}^n \frac{d^3\bp_i}{(2\pi)^3 2E_i}\ket{\bp_1,\ldots,\bp_n} \bra{\bp_1,\ldots,\bp_n} =1
\eeq
and $p'=\sum_i p_i$ is the total momentum of the state $X$. 

The momentum conserving $\delta$-functions in \eq{sig1} constrain $p'$ and thus allow only a subset of the states $X$ in the completeness sum \eq{comp}. The restriction on ${\bp'}_\perp$ is avoided by summing over all transverse momenta $\bqt$ of the virtual photon. 
As mentioned in section 2, large values of $\bqt$ do not contribute significantly to the sum due to the suppression provided by the GPD's. On the other hand, 
${p'}^+$ and ${p'}^-$ are fixed by a measurement of $x_B$ and $x_F$. Hence we need to incorporate the corresponding $\delta$-functions in the (hermitian conjugate) matrix element:
\beqa\label{shift}
\bra{N(p)}\bar\psi_d(0)\gamma^+\gamma_5\,\psi_u(y_2) \ket{X(p')}\,2 (2\pi)^2 \delta(p^+-q^+-{p'}^+) \delta(k^-+p^--q^--{p'}^-) \hspace{.5cm}\nn\\
= \halft \int dy_3^+\,dy_3^-\, \bra{N(p)}\bar\psi_d(y_3)\gamma^+\gamma_5\,\psi_u(y_2+y_3) \ket{X(p')} \exp\left[iy_3\cdot(k-q)\right]
\eeqa
where $\bs{y}_{3\perp}=0$ and we used translation invariance in the matrix element. The full completeness sum is allowed using the matrix element in \eq{shift}, since states $X$ which do not conserve momentum will not contribute after the $y_3$-integrations. The inclusive cross section \eq{sig1} will thus be given by the multiparton distribution shown in \fig{fig3},

\EPSFIGURE[h]{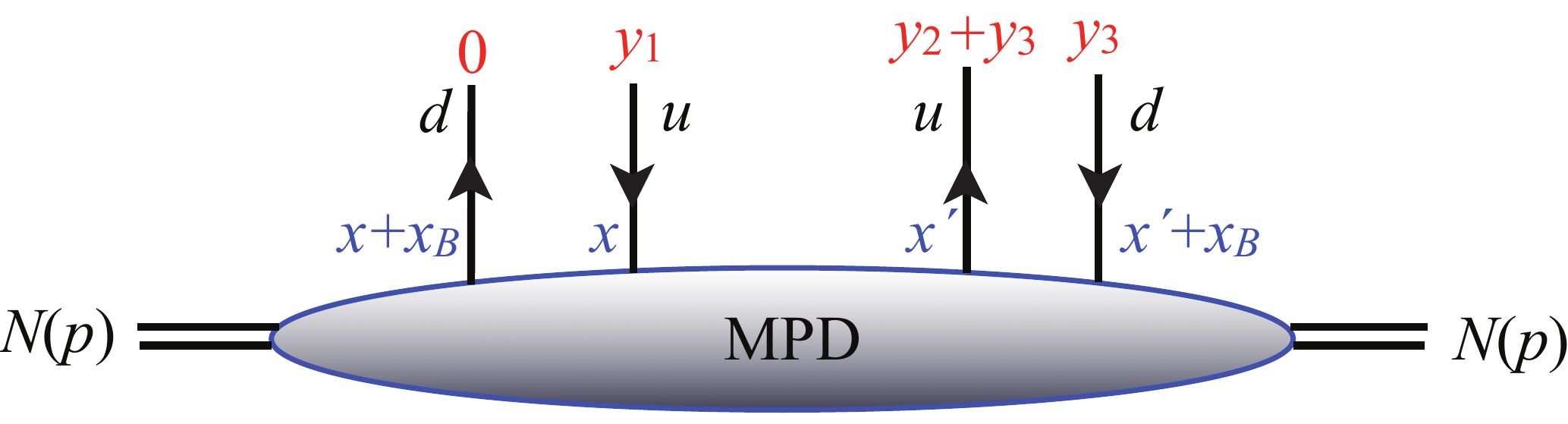,width=0.9\columnwidth}{Pictorial representation of the forward multiparton distribution $f_{d\bar u/p}(x_B,x_M;x,x')$ given in Eq.~\eq{dpdf}.
\label{fig3}}

\beqa\label{dpdf}
f_{d\bar u/p}(x_B,x_M;x,x') = \hspace{10cm}\\
=\frac{1}{4(4\pi)^3}
\int  dy_1^- dy_2^- dy_3^- dy_3^+ \exp\left\{\halft i\left[-y_1^-l_1^+ +y_2^-{l_1^+}'- y_3^-q^+ +y_3^+ x_Mp^-
\right]\right\} \nn\\ \nn\\
\times  \bra{N(p)}\bar\psi_d(y_3)\gamma^+\gamma_5\,\psi_u(y_2+y_3)\, \bar\psi_u(y_1)\gamma^+\gamma_5\,\psi_d(0) \ket{N(p)}_{y_{i\perp}=0;\ y_1^+=y_2^+=0}\nn
\eeqa
where $x_B=q^+/p^+$, $x=l_1^+/p^+$, $x'={l_1'}^+/p^+$ and the scaled `$-$' momentum transferred to the inclusive system is denoted $x_M=k^-(1-x_F)/p^-$. The inclusive mass $M_X$ is given by $x_M$ as
\beq\label{xmass2}
M_X^2 = m_N^2(1-x_B)(1+x_M)-\bs{q}_\perp^2 
\eeq
The kinematic range of $x_M$ in the BB limit is thus
\beq\label{xMrange}
\frac{x_B+\bs{q}_\perp^2/m_N^2}{1-x_B} \leq x_M \le \infty
\eeq

The shift by $y_3$ introduced in \eq{shift} between the fields in the matrix elements of $T$ and in $T^\dag$ is conjugate to the momentum transfer $k-q$ between the hard vertex and the target. The MPD \eq{dpdf} differs from the higher twist distributions discussed by 
Jaffe \cite{Jaffe:1983hp} through its dependence on the finite LF time difference $y_3^+$ between the fields. 
Also the standard leading twist PDF's are evaluated at $y_3^+ =0$ since the inclusive system $X$ carries an asymptotically large `$-$' momentum in the Bjorken limit.

In the BB limit the the momentum transfer $x_M p^-=k^-(1-x_F) = (l_1-l_2)^- \sim 1/y_3^+$ to the target is kept finite. We may, however, consider the case where this transfer is nevertheless large compared to $p^-$. Since the quark lines $l_1$ and $l_2$ connect to the non-perturbative matrix element their virtualities $l_1^2,\ l_2^2$ should remain limited, thus $l_1^+ \sim \lqcd^2/l_1^-$ and similarly for $l_2^+$. Given that $l_2^+ - l_1^+ = q^+ = x_B p^+$ is fixed, either $l_1^-$ or $-l_2^-$ may be large, while the other has to be of \order{p^-}. If $l_1^-$ is large then $l_1^+ \simeq 0$ and $l_2^+ \simeq q^+$. In this kinematics the $u$-quark will hadronize independently into a final state jet \cite{Berger:1979du} and the target MPD \eq{dpdf} should reduce to a $d$-quark PDF. We demonstrate this in  Appendix~B.

Using the expression \eq{amp9} of the scattering amplitude in the cross section \eq{sig1} and incorporating the longitudinal $\delta$-function constraints in $T^\dag$ as in \eq{shift} the Drell-Yan cross section in the BB limit can be expressed in terms of the double parton distribution \eq{dpdf} as
\beqa\label{sig2}
\frac{d\sigma(\pi^+ N\to\gamma^*_L X)}{dM_X^2} &=& \frac{2(eg^2C_F)^2}
{Q^2 s^2(1-x_B)N_c}
\int dx\,dx'\, C(x_B,x) C^*(x_B,x')\, f_{d\bar u/p}(x_B,x_M;x,x')\nn\\
\eeqa
Including the muon pair observables as in \eq{ampmu} the differential cross section becomes
\beq\label{sig3}
\frac{d\sigma(\pi^+ N\to\mu\mu X)}{dx_B\,d\Omega_{\mu\mu}\,dM_X^2}
=\frac{2\sin^2\theta}{(4\pi)^3 x_B} \frac{d\sigma(\pi^+ N\to\gamma^*_L X)}{dM_X^2}
\eeq

\section{Photon helicity for pion and proton induced Drell-Yan processes}

\EPSFIGURE[r]{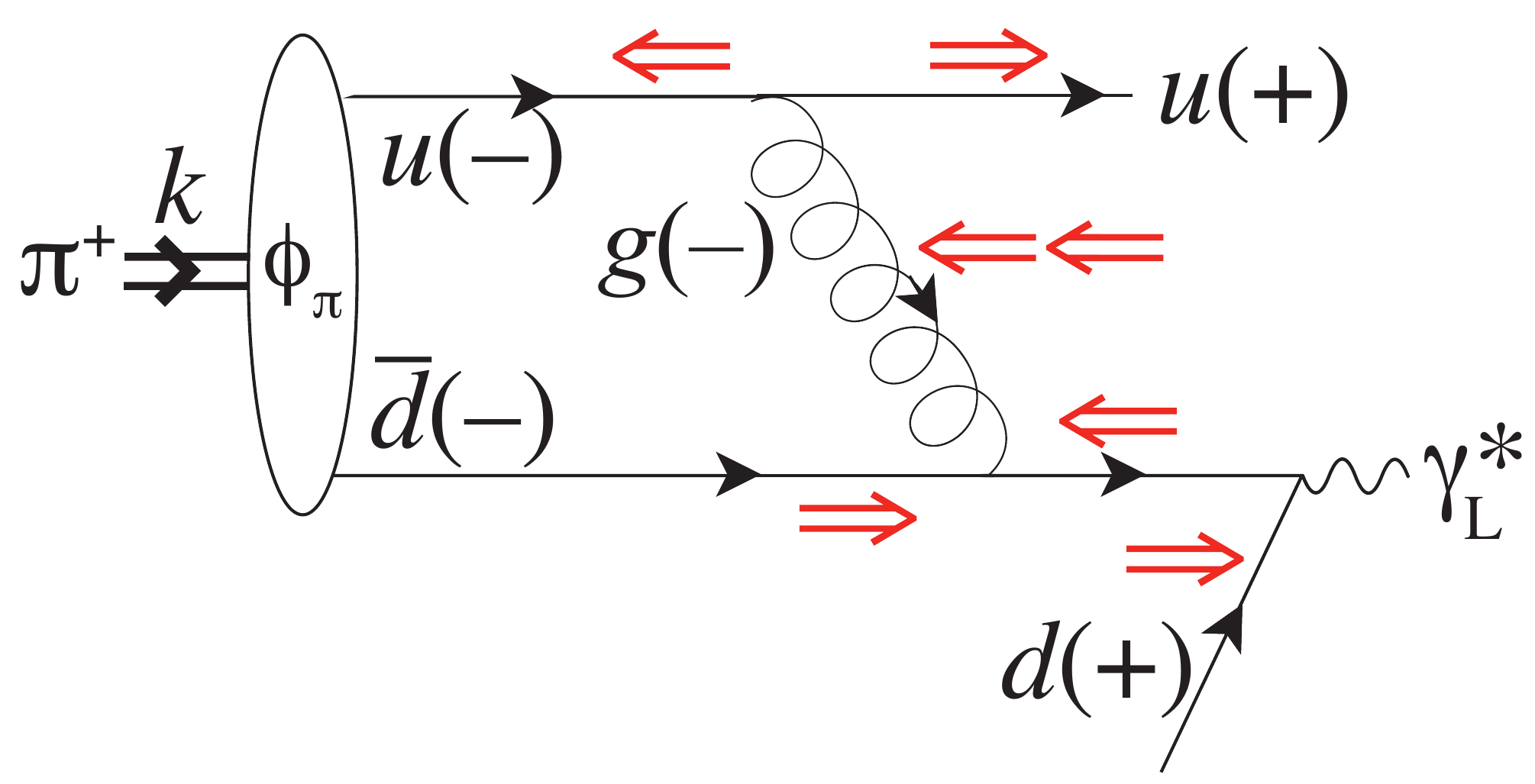,width=0.5\columnwidth}{The double red arrows indicate the spin directions of the particles. All momenta are in the $\pm z$-direction as shown in the parentheses. The $u$-quark propagates into the GPD 
(not shown)
while the $d$-quark propagates out of it.
\label{fig4}}

The fact that the photon is coherent with an entire pion Fock state makes it natural that it also carries the helicity $(\lambda=0)$ of the pion. On the other hand, this argument does not suffice to determine the photon polarization in $pN \to \gamma^*X$, since there is a minimal helicity flip $|\Delta\lambda|=\halft$ for both transverse and longitudinal photons. 
The photon was found to be transversely polarized in a calculation of the exclusive $\bar p N \to \gamma^*\pi$ process \cite{Pire:2005ax}.
As we shall see, the same result is obtained for the proton induced inclusive Drell-Yan process in the BB limit. 

The helicity systematics follows in a straightforward way from three facts:\\

\noindent
(i) The hard interactions conserve quark helicity up to corrections of \order{m_q/Q};\\
\noindent
(ii) Since all transverse momenta $q_\perp$ are limited, orbital contributions $L_z \sim \morder{q_\perp/Q}$;\\
\noindent
(iii) Angular momentum $J_z \simeq S_z$ is conserved. \\

The helicities are then obtained by simple addition. In \fig{fig4} the $J_z$ components are indicated by double arrows 
in the case $x>0$ where the $u$-quark propagates (with positive energy) into the GPD while the $d$-quark propagates out of it.
They follow from helicity conservation for the quark lines, taking into account the direction of longitudinal momentum of the particles (indicated by $\pm$ for motion in the $\pm z$ direction). {\it E.g.,} the dominant spin component $S_z = -1$ of the gluon may be verified directly by expressing its propagator as a sum over helicities 
as in \eq{helsum}.
The quark propagators can analogously be expanded in terms of the spinors \eq{lfspinors}.

\EPSFIGURE[tb]{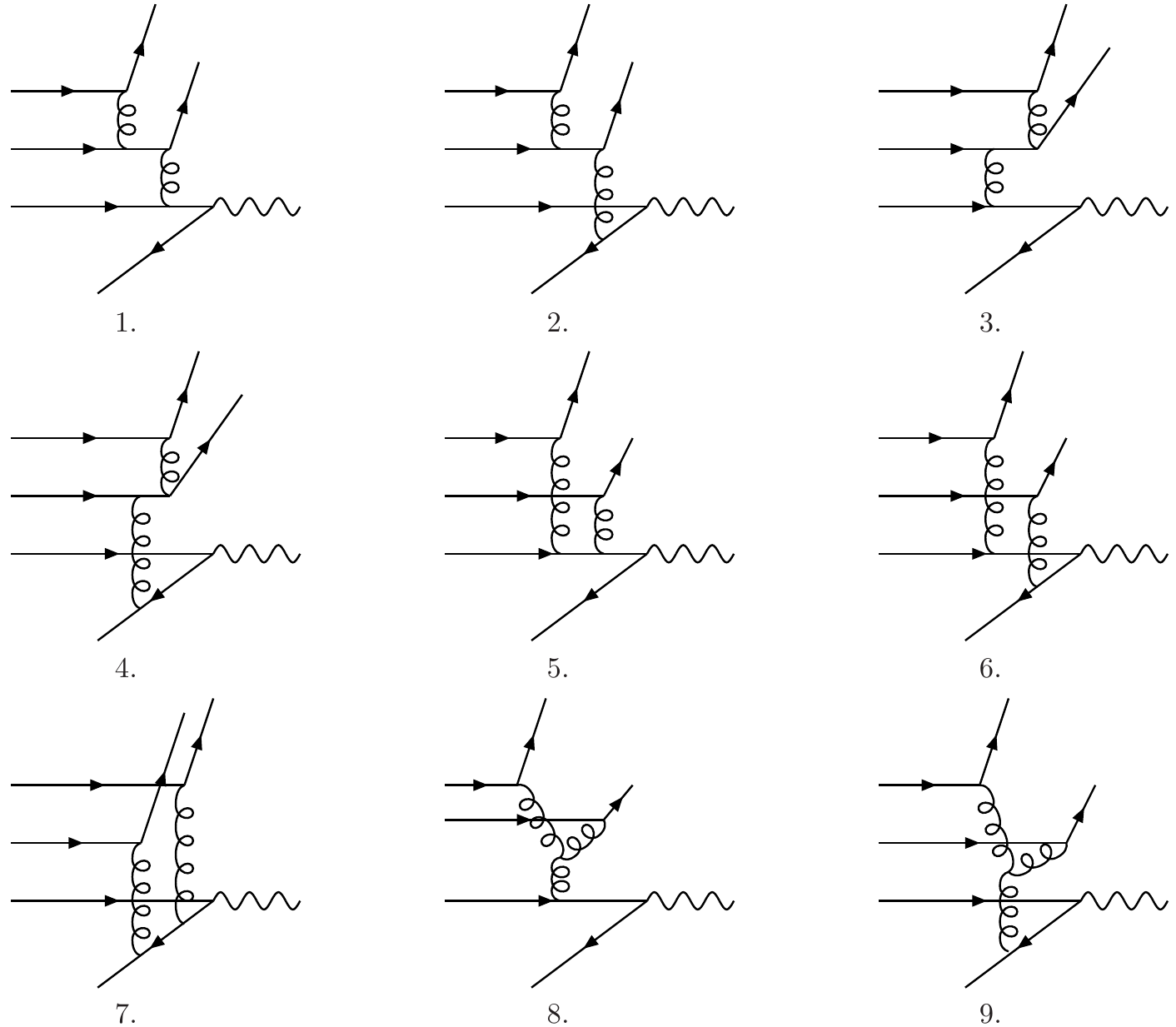,width=0.72\columnwidth}{Nine topologically different subprocess diagrams in $p\bar p \to \gamma^*(x_F) + X$ at  large $x_F$. 
\label{fig5}}

The photon polarization in the case of a $J=\halft$ beam particle, such as a nucleon, may be similarly deduced. The analysis is especially simple if we model
the nucleon as scalar diquark-quark bound state. Setting $S_z(u)=0$ in \fig{fig4} gives $S_z(g)=0$ and consequently 
$S_z(\gamma^*)= 1$.

We have calculated the relevant diagrams (Fig.~\ref{fig5}) also for three-quark Fock states of a nucleon 
(see \cite{Pire:2005ax} for a corresponding calculation of the exclusive $\bar p N \to \gamma^*\pi$ process).
Ignoring flavor, the hard subprocess involves nine topologically different diagrams at tree level. 
See Fig.~\ref{fig6} for momentum and helicity labels. We parametrize the momenta in analogy to \eq{kmom}, \eq{lmom} as
\beqa \label{eq:momdef}
 k_1&=&\left(0^+,z_1 k^-,\kveco\right)\nn\\
 k_2&=&\left(0^+,z_2 k^-,\kvect\right)\nn\\
 k_3&=&\left(0^+,(1-z_1-z_2) k^-,-\kveco-\kvect\right)\nn\\
 \ell_1&=&\left(x_1 p^+,0^- ,\lveco\right)\nn\\
\ell_2&=&\left(x_2 p^+,0^- ,\lvect\right)\nn\\
\ell_3&=&\left(-(x_B+x_1+x_2) p^+,0^- ,-\lveco-\lvect\right)\nn\\
 q &=&\left(x_B p^+, x_F k^- ,\notvec\right)
\eeqa
where all transverse momenta ($\sim \Lambda_{QCD}$) are negligible at leading order in $1/Q$ when evaluating the hard subprocess.

We illustrate the method by calculating the diagram of Fig.~\ref{fig6} in the BB limit in Feynman gauge. We suppress all color indices and use the light front spinors of \eq{lfspinors} for the proton valence quarks.
Helicity conservation along the quark lines fixes $t_{1,2,3} = s_{1,2,3}$, leaving $s_1,s_2,s_3,$ and $\la$ as parameters. We find that four helicity amplitudes ${\cal M}^\la_{s_1,s_2,s_3}$ receive  contributions at leading order (${\cal M}^\la_{s_1,s_2,s_3} \sim 1/Q^3 $):
\beqa
 & & {\cal M}^{+1}_{+,-,+} \simeq {\cal M}^{+1}_{-,+,+} \simeq   
\frac{4 \sqrt{2} e_q g^4 \sqrt{ x_2 (x_1+x_2+x_B)  z_2 (1-z_1-z_2)} }{(p^+k^-)^{3/2} \sqrt{x_1z_1} (x_1+x_2)^2  (z_1+z_2)^2} \nn \\
&\simeq& {\cal M}^{-1}_{-,+,-} \simeq {\cal M}^{-1}_{+,-,-} 
\eeqa
where the lower indices $\pm$ stand for $\pm \halft$ and $e_q$ is the quark charge at the photon vertex. The virtual photon is transverse in the leading amplitudes. This result can also be deduced by using helicity and angular momentum conservation as explained above. The z-components of spins are marked with red double arrows in Fig.~\ref{fig6} for the amplitude ${\cal M}^{+1}_{+,-,+}$. Notice that the spin directions match with those of the scalar diquark model above.

\EPSFIGURE[hr]{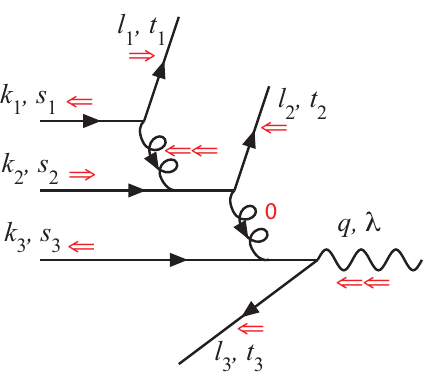,width=.45\columnwidth}{
Definitions of momenta $(k_i,\ell_i)$ and helicities $(s_i,t_i)$ in diagram~1 of \fig{fig5}. The red arrows indicate $S_z$ in units of $\halft\hbar$ when $s_1=-s_2=s_3=\halft$ and the photon has helicity $\lambda=1$. The second gluon carries $S_z=0$.
\label{fig6}}

We have checked by explicit calculation that the above reasoning applies similarly to all the diagrams of Fig.~\ref{fig5}. 
We find that the following helicity combinations are leading (fixing $\la=+1$) :
\beqa \label{eq:contlist}
 \textrm{Diagram:} & \quad & \textrm{Helicities:}\nn\\
 \textrm{1,4,8,9} & \quad & (s_1,s_2,s_3) = (+,-,+),\ (-,+,+)\nn\\
 \textrm{5} & \quad & (s_1,s_2,s_3) = (-,+,+)\nn\\
 \textrm{6} & \quad & (s_1,s_2,s_3) = (+,+,-)\nn\\
 \textrm{7} & \quad & (s_1,s_2,s_3) = (+,-,+)
\eeqa
\noindent Diagrams~2 and~3 do not contribute in Feynman gauge. Contributions with $\la=-1$ are obtained from \eq{eq:contlist} by flipping all quark helicities, and amplitudes with $\la=0$ are subleading as expected. 

Interestingly, the $(+,+,-)$ amplitude gets a contribution only from diagram~6
of Fig.~\ref{fig5}.
This can be understood in terms of helicity conservation since the rules (i)-(iii) imply that for all other diagrams of Fig.~\ref{fig5} except diagram~6 either a gluon has $|S_z|=2$ or a quark has $|S_z|=3/2$. Note that diagram~6 is gauge invariant at leading order in $1/Q$ since both gluons are transverse.

It is also straightforward to check that amplitudes where a pion (proton) projectile produces a transverse (longitudinal) virtual photon are suppressed by orbital factors of 
$|\boldsymbol{\ell}_{1,2\perp}|/Q$ or  $|\boldsymbol{k}_{1,2\perp}|/Q$ as expected from rule (ii) above. In the suppressed amplitudes the projectile wave function is strongly weighted at the endpoints $z=0,1$. {\it E.g.,} the pion distribution amplitude is weighted by $1/z^2$ for $z\to 0$ in the transverse photon amplitude, instead of by $1/z$ as in \eq{Cdef}. This endangers the convergence of the $z$-integral and is related to the difficulty of proving factorization for $\gamma^*_T N \to \pi N$ \cite{Collins:1996fb}.

\section{Discussion}

The remarkable equality of the inclusive $\gamma^* N \to X$ DIS data at high $Q^2$ with the exclusive resonance contributions at lower $Q^2$ \cite{duality05} points to an intriguing relation between inclusive and exclusive processes. This Bloom-Gilman duality indicates that two mathematical limits simultaneously describe the data: $Q^2\to \infty$ at fixed $x_B$ and at fixed hadronic (resonance) mass $W$. The latter limit is equivalent to the BB limit of keeping $Q^2(1-x_B)$ fixed, some aspects of which we have explored in this paper.

We focussed on the Drell-Yan process $\pi N \to \gamma^* X$ at fixed $Q^2(1-x_F)$ for the virtual photon. The $x_F\to 1$ limit forces a single quark to carry nearly all the pion momentum. Knowing the ``end-point'' behaviour of the pion wave function is important in the analysis of
its exclusive form factor. The DY data \cite{Anderson:1979xx} shows that the photon is longitudinally polarized at high $x_F$, implying that the annihilating quark(s) are far off-shell. This dynamics was studied by Berger and Brodsky \cite{Berger:1979du}, who showed that the hard subprocess is different from the standard $q\bar q \to \gamma^*$ one and, as in exclusive processes, is coherent with the entire pion Fock state.

The theoretical and experimental evidence of the dominance of a hard subprocess which is suppressed (\ie, of higher twist) in the usual limit ($Q^2 \to \infty$ at fixed $x_F$) raised the question whether one can define another limit where this subprocess would be a leading contribution and could be factorized from the soft dynamics. As we discussed in section 1, the requirement of coherence naturally selects the BB limit of fixed $Q^2(1-x_F)$. Since the mass of the inclusive system $X$ is fixed in this limit the existing factorization proofs for deeply exclusive meson production $\gamma^* N \to \pi N$ \cite{Collins:1996fb} apply. The derivation is valid for any fixed mass state $X$ in $\pi N \to\gamma^* X$, which builds the inclusive Drell-Yan process. The elastic process $\pi N \to \gamma^* N$ was already considered in \cite{Berger:2001zn}. For $X \neq N$ the soft dynamics involves ``Transition Generalized Parton Distributions'', which have also been studied previously 
\cite{Frankfurt:1999fp,Pire:2005ax}. 

We found that  $\pi N \to \gamma^* X$ differs essentially from $p N \to \gamma^* X$ in the BB limit since the photon is longitudinally polarized in $\pi N$ and transversally in $pN$. Hence the transition from the Bj to the BB limit can be directly observed in $\pi N$ through the change of photon polarization with increasing $x_F$. In $pN$ there is (as expected) no evidence for a change of polarization from transverse to longitudinal, and the two limits may coexist in a range of $x_F$ as suggested by duality \cite{duality05}. This may be connected to the observation that 
the azimuthal distribution of the muon pair in $p N \to \mu^+\mu^- X$ is consistent with the standard QCD analysis \cite{Zhu:2006gx} while this is not the case for $\pi N \to \mu^+\mu^- X$ \cite{Falciano:1986wk}.

The BB limit seems appropriate for understanding the large single spin asymmetries (SSA) observed in hadron scattering at high $x_F$ and large transverse momenta \cite{Hoyer:2006hu}. The SSA requires quark helicity flip and a dynamical phase, both of which are suppressed in hard subprocesses. The coherence between partons of high and low $x$ allows the helicity flip and phase to occur in the soft part of the amplitude where they are not suppressed.

The multiparton distribution
\eq{dpdf} that describes
the soft dynamics of the Drell-Yan process in the BB limit is a forward target matrix element with four quark fields, which would be of higher twist in the usual Bj limit of fixed $x_F$. It differs from the MPD's studied by Jaffe \cite{Jaffe:1983hp} in that the two pairs of quark fields originating from the amplitude and its complex conjugate are evaluated at a finite LF time difference $y_3^+$. This is a consequence of the finite (in the target rest frame) `minus' momentum $(1+x_M) p^-$ of the inclusive system and means that the ordering of the quark fields in the matrix element is significant. For $x_M \to \infty$ we demonstrated that the main contribution to the MPD is from a contraction of a pair of quark fields, which reduces the MPD to a standard leading twist parton distribution.

\vspace{.5cm}

\noindent {\bf Acknowledgments}

\vspace{.2cm}

We wish to thank Stan Brodsky, Markus Diehl and St\'ephane Peign\'e 
for helpful discussions. PH is grateful for a travel grant from the Magnus Ehrnrooth foundation. 
MJ has been supported in part by the Marie Curie Excellence Grant under contract MEXT-CT-2004-013510.

\appendix

\vspace{1cm}
\centerline{\Large \bf Appendix}

\section{The virtual photon and pion wave functions}

We use LF polarization vectors \cite{Brodsky:1997de} for the virtual photon with helicity $\lambda$,
\beq\label{polvec}
\veps_\lambda(q) = e_\lambda-\frac{e_\lambda \cdot q}{q^+} n
\eeq
with
\beqa
e_{\pm 1}&=& -\inv{\sqrt{2}}(0,0,\pm 1,i) \hspace{2cm} n = (0,2,0,0)\nn\\ \\
e_0(q) &=& -\frac{iq}{\sqrt{q^2}}\hspace{3.7cm} \tilde{n} = (2,0,0,0)\nn
\eeqa
which satisfy
\beq\label{helsum}
\sum_{\lambda=\pm 1, 0} \veps_\lambda^\mu(k)\, {\veps_\lambda^\nu(k)}^* = -g^{\mu\nu} + \frac{k^\mu k^\nu}{k^2}
\eeq

The pion valence state is defined by its wave function $\Psi_{\bk}$,
\beq
\ket{\pi(\bk)} = \sum_{s,A}\frac{(-1)^{s-\half}}{\sqrt{2N_c}}\int \frac{d^3\bk_1\, \Psi_{\bk}(\bk_1)}{\sqrt{4|\bk_1||\bk_2|}(2\pi)^3} b^\dag_{s,A}(\bk_1) d^\dag_{-s,A}(\bk_2)\ket{0}
\eeq
where $\bk_2 = \bk-\bk_1$. The sum is over the helicity $s=\pm \halft$ and color $A$ ($N_c=3$). With
\beq
\acom{b_{s,A}(\bk_1)}{b^\dag_{s',A'}(\bk_1')} = (2\pi)^3 2|\bk_1| \delta^3(\bk_1-\bk_1') \delta_{ss'} \delta_{AA'}
\eeq
this gives the normalization
\beq
\langle{\pi(\bk')}\ket{\pi(\bk)} = (2\pi)^3 2|\bk| \delta^3(\bk-\bk') \int \frac{d^3\bk_1}{(2\pi)^3 2|\bk|} |\Psi_{\bk}(\bk_1)|^2
\eeq
Due to the large virtuality of the gluon propagator $q_1$ in Fig.~\ref{fig1} only transversally compact valence Fock states contribute at leading order, involving the pion distribution amplitude
\beq\label{distamp}
\phi_\pi(z) \equiv \int \frac{d^2\bk_{1\perp}}{2(2\pi)^3}\Psi_{\bk}(\bk_1)
\eeq
where $k_1^- = zk^-$ as in \eq{kmom}. With $f_\pi \simeq 93$ MeV the normalization is
\beq
\int_0^1 dz\,\phi_\pi(z) = -\frac{f_\pi}{2\sqrt{N_c}}
\eeq

Since we took the pion to move along the $-z$ direction in \eq{momdef} we define the LF spinors \cite{Brodsky:1997de} in the quark operator
\beq
\psi(y) = \int \frac{d^3\bp}{2|\bp|(2\pi)^3}\sum_s \left[b_s(\bp)u_s(\bp)e^{-ip\cdot y} + d^\dag_s(\bp)v_s(\bp)e^{ip\cdot y}\right]
\eeq
to be well-defined for $k^+=0$,
\beqa\label{lfspinors}
 u(p,s) &=& \frac{1}{\sqrt{p^-}} \aslash{p}\ \chi(s)\nn\\
 v(p,s) &=& \frac{1}{\sqrt{p^-}} \aslash{p}\ \chi(-s)
\eeqa
where $\chi(\halft)=(1,0,1,0)^T/\sqrt{2}$ and $\chi(-\halft)=(0,1,0,-1)^T/\sqrt{2}$.

At lowest order the Bethe-Salpeter wave function of the pion is
\beqa
\Phi_{\bk}^{\alpha\beta}(y_1,y_2) \equiv \bra{0}\bar\psi_\beta^B(y_2)\psi_\alpha^A(y_1) \ket{\pi(\bk)}\hspace{8.5cm}\\
= \frac{\delta_{AB}}{\sqrt{2N_c}} \sum_s (-1)^{s-\half}\int \frac{d^3\bk_1\, \Psi_{\bk}(\bk_1)}{\sqrt{4|\bk_1||\bk_2|}(2\pi)^3} \bar v_\beta(\bk_2,-s) u_\alpha(\bk_1,s)\exp(-ik_1\cdot y_1-ik_2 \cdot y_2)\nn
\eeqa
For the LF spinors defined in \eq{lfspinors},
\beqa
\sum_s (-1)^{s-\half}\bar v_\beta(\bk_2,-s) u_\alpha(\bk_1,s) &=& \inv{\sqrt{k_1^- k_2^-}} \sum_s(-1)^{s-\half} (\slashed{k}_1)_{\alpha\alpha'}\chi_{\alpha'}(s) 
{(\gamma^0\chi)}_{\beta'}^\dag(s) 
(\slashed{k}_2)_{\beta'\beta}\nn\\
&=& \frac{-1}{2\sqrt{k_1^- k_2^-}}[\slashed{k}_1 \slashed{\tilde n}\gamma_5 \slashed{k}_2]_{\alpha\beta} \simeq \inv{2} \sqrt{k_1^- k_2^-}\,[\slashed{n}\gamma_5]_{\alpha\beta}
\eeqa
where the last equality applies in the limit $k^- \to\infty$, using $\slashed{k}_i \simeq \halft k_i^- \slashed{n}$. Thus
\beq\label{BSampl}
\Phi_{k^-\to\infty}^{\alpha\beta}(y_1,y_2) = \frac{\delta_{AB}}{2\sqrt{2N_c}} [\slashed{n}\gamma_5]_{\alpha\beta}\, k^- \int \frac{dz\,d^2\bk_{\perp 1}}{16\pi^3} \, 
\Psi_{\bk}(k_1^z=-\halft zk^-,\bk_{\perp 1}) 
\exp(-ik_1\cdot y_1-ik_2 \cdot y_2)
\eeq
With $k_1^+= k_2^+ = 0$ as in \eq{kmom} the BS wave function $\Phi_{k^-\to\infty}^{\alpha\beta}(y_1,y_2)$ is independent of $y_1^-, y_2^-$. Thus the wave function is the same whether evaluated at $y_1^-= y_2^-$ (the LF wave function for a particle moving in the $-z$ direction) or at equal ordinary time, $y_1^- - y_2^- = y_2^+ - y_1^+$. This expresses the well-known fact that
equal time wave functions are the same as LF wave functions
in the infinite momentum frame.

In the BB limit we have 
$y_{i\perp}=0$ 
($i=1,2$), hence the pion enters through its distribution amplitude \eq{distamp}. 
In analogy to \eq{effrules} we can write an effective Feynman rule for the pion 
\beqa\label{pirule}
 u_\beta(k_1) 
\bar v_\alpha(k_2) \to \int \frac{dk_1^-}{2\pi}\frac{dy_1^+}{2} \exp\left( \halft ik_1^-y_1^+ \right)\bra{0}\bar\psi_\beta^B(0)\psi_\alpha^A(y_1)\ket{\pi(\bk)}_{y_1^- = y_{1\perp}=0}
\hspace{2.5cm} \\
 = \int \frac{dk_1^-}{2\pi}\frac{dy_1^+}{2}\Phi_{\bk}^{\alpha\beta}(y_1^+,0) \exp\left( \halft ik_1^-y_1^+ \right)\ \ \substack{\phantom{abc} \\ \simeq \\ {\scriptscriptstyle k^-\to \infty}}\ \ \frac{\delta_{AB}}{2\sqrt{2N_c}} [\slashed{n}\gamma_5]_{\alpha\beta}\, k^- \int dz \, \phi_\pi(z) \nn
\eeqa
which will be multiplied by $k_1^- = zk^-$-dependent propagators, \cf \eq{ellvirt}.

\section{The limit of large $x_M p^- = k^-(1-x_F)$}

In the BB limit of the DY process $\pi^+ N \to \gamma^* X$ the virtual photon carries nearly all ($x_F \to 1$) of the pion momentum, leaving a finite (in the target rest frame) transfer $x_M p^- = k^-(1-x_F)$ to the target system. In this Appendix we consider the case $x_M \gg 1$, when the momentum transfer $(l_1-l_2)^-$ is large compared to $p^-$. Since the quark lines $l_1,\ l_2$ attach to the non-perturbative GPD their virtualities $l_i^2 = l_i^+ l_i^--l_{i\perp}^2$ should remain of \order{\lqcd^2}. Given that $(l_2-l_1)^+ = x_B p^+$ is fixed either $l_1^-$ or $-l_2^-$, but not both, can grow large. When $l_1^-$ is large the $u(l_1)$-quark hadronizes independently of the target remnants, similarly to the struck quark in the standard Bjorken limit of the DY process. Here we show that the target MPD in the expression \eq{sig2} of the DY cross section reduces to the $d$-quark PDF as $x_M p^- \simeq l_1^- \to \infty$, and to the $\bar u$-quark PDF as $x_M p^- \simeq -l_2^- \to \infty$.

\vspace{.3cm}
\noindent{\it (i) $f_{d\bar u/p} \to f_{d/p}$ as $x_M\to \infty$ with $x_Mp^- \simeq l_1^-$}
\vspace{.2cm}

According to the expression \eq{dpdf} of the MPD the LF time difference between the quark fields  $y_3^+ \sim 1/x_M p^- \to 0$ as $x_M \to \infty$. The dynamics then becomes light-cone dominated and contractions of the $u$- or $d$-quark fields in the MPD dominate. A contraction turns the MPD into a standard PDF in which the field ordering is irrelevant, \ie, only one contraction is possible.

The contribution where the $u$-quark has large momentum and is treated as a free final state particle
$(l_1^- \to \infty,\ l_1^+ \sim \lqcd^2/l_1^-)$ 
corresponds to contracting the $u$-fields in $f_{d\bar u/p}$ of  \eq{dpdf}. Using
\beq\label{contr1}
\bra{0}\psi(x)\bar\psi(0)\ket{0} = \left.\int\frac{d^3\bs{l}}{(2\pi)^3 2|\bs{l}|}e^{-il\cdot x} \slashed{l}\,\right|_{l^0=|\bs{l}|} = 
\left.\int\frac{dl^+\,d^2\bs{l}_\perp}{(2\pi)^3 2l^+}\theta(l^+)\,e^{-il\cdot x} \slashed{l}\,\right|_{l^-=l_\perp^2/l^+}
\eeq
we have
\beqa\label{fdu1}
f_{d\bar u/p}&=&\frac{1}{4(4\pi)^3}
\int  dy_1^- dy_2^- dy_3^- dy_3^+ \exp\left\{\halft i\left[-y_1^-l_1^+ +y_2^-{l_1^+}'- y_3^-q^+ +y_3^+x_M p^-\right]\right\} \nn\\
&\times&\int\frac{dl^+\,d^2\bs{l}_\perp}{(2\pi)^3 2l^+}\theta(l^+)\,2l^+ \bra{N(p)}\bar\psi_d(y_3)\gamma^+\,\psi_d(0) \ket{N(p)} \nn\\
&\times& \exp\left[-\halft il^+(y_2^-+y_3^--y_1^-)-\halft i \frac{l_\perp^2}{l^+}y_3^+\right]
\eeqa
Integrating over the transverse momenta,
\beq
\int d^2\bs{l}_\perp \exp\left[- i\, \frac{l_\perp^2}{2l^+}\, y_3^+\right] = -\frac{2\pi il^+}{y_3^+-\ieps}
\eeq
Given
the small effective range of $y_3^+ \sim 1/(k^-(1-x_F))$
we may neglect the $y_3^+$-dependence of the matrix element in \eq{fdu1} (which is regular on the light-cone). We get
\beq
\int dy_3^+ \frac{\exp[\halft i y_3^+ k^-(1-x_F)]}{y_3^+-\ieps} = 2\pi i
\eeq
The integrals over $y_1^-$ and $y_2^-$ give $\delta$-functions constraining $l_1^+ = {l_1'}^+$ and $l^+ = l_1^+$. Noting that $l_2^+ = q^+ + l_1^+$ we have
\beq\label{fdu2}
f_{d\bar u/p} = \delta(l_1^+ - {l_1'}^+)\, \frac{l_1^+}{4\pi}\,\theta(l_1^+) f_{d/p}(l_2^+/p^+)
\eeq
where the $d$-quark PDF is
\beq
f_{d/p}(l_2^+/p^+) = \frac{1}{8\pi}\int dy_3^-\,\exp(-\halft i y_3^-l_2^+)\, \bra{N(p)}\bar\psi_d(y_3^-)\gamma^+\,\psi_d(0) \ket{N(p)}
\eeq

The dominant contribution to the cross section \eq{sig2} comes from the region of low $l_1^+ = xp^+$ due to $|C(x_B,x)|^2 \propto 1/x^2$ (the other singularity at $x = -x_B$ is outside the kinematic region since $l_1^+ > 0$). 
This is consistent with $l_1^+ \propto 1/l_1^- \to 0$ at large $l_1^-$, which ensures that the independently hadronizing $u$-quark is nearly on-shell.
Thus
\beq
C(x_B,x) \simeq \frac{e_d}{x}\int \frac{dz}{z}\phi_\pi(z)
\eeq
Using this and the expression \eq{fdu2} for the MPD in the cross section \eq{sig2} we get 
\beq\label{sig4}
\frac{d\sigma(\pi^+ N\to\gamma^*_L X)}{dM_X^2} = \frac{(ee_dg^2C_F)^2}
{Q^2 s^2(1-x_B)N_c}
\int \frac{dl_1^+}{2\pi l_1^+}\,\theta(l_1^+)\, \left(\int \frac{dz}{z}\phi_\pi(z)\right)^2\, f_{d/p}(l_2^+/p^+)
\eeq

\vspace{.3cm}
\noindent{\it (ii) Comparison with standard factorization}
\vspace{.2cm}

We should compare the cross section \eq{sig4} with a calculation \cite{Berger:1979du} where $\sigma(\pi^+ + N \to u + \gamma_L^*(q) + X)$ is expressed in terms of the hard subprocess cross section $\hat\sigma(\pi^+(k)+ d(l_2) \to u(l_1)+\gamma_L^*(q))$ convoluted with $f_{d/p}(l_2^+/p^+)$. Parametrizing the momenta as
\beqa
l_1 &=& (l_1^+, l_1^-,\bs{l}_\perp)\nn\\
q &=& (q^+,q^-,-\bs{l}_\perp)\\
l_2 &=& (l_2^+,0^-,\boldsymbol{0}_\perp)\nn
\eeqa
we have $l_1^- \simeq x_M p^-$, $l_1^+ = l_2^+-q^+$, $l_\perp^2 = l_1^+ l_1^-$
and $Q^2 \simeq q^+q^-$. The subprocess amplitude is in the BB limit
\beq
\hat T(\pi^+ d \to u+\gamma_L^*) = \mp\frac{4iee_dg^2C_F}{Q\sqrt{2N_c}}\,\sqrt{\frac{l_2^+}{l_1^+}}
 \int\frac{dz}{z}\phi_\pi(z)
\eeq
where the signs correspond to the helicity $\pm \halft$ of the $u$-quark. The subprocess cross section
\beqa\label{sig5}
\hat\sigma(\pi^+ d \to u+\gamma_L^*) &=& \frac{1}{16\pi\hat s}\int dl_1^+\,\theta(l_1^+)\, dq^- dl_\perp^2 \delta(l_1^+ l_1^--l_\perp^2)
\,\delta(q^+q^- -Q^2)|\,\hat T\,|^2 \nn \\
&=& \frac{(ee_dg^2C_F)^2}{Q^4 N_c}
\int \frac{dl_1^+}{2\pi l_1^+}\,\theta(l_1^+)\, \left(\int \frac{dz}{z}\phi_\pi(z)\right)^2
\eeqa
gives the full cross section 
\beq
\sigma(\pi^+ N \to u + \gamma_L^* + X)=\int \frac{dl_2^+}{p^+}\hat\sigma(\pi^+ d \to u + \gamma_L^*)\, f_{d/p}(\frac{l_2^+}{p^+})
\eeq
Using $dl_2^+ = dq^+ = dM_X^2\,Q^2/[sk^-(1-x_B)]$ this expression agrees with \eq{sig4} derived from the MPD.

\vspace{.3cm}
\noindent{\it (iii) $f_{d\bar u/p} \to f_{\bar u/p}$ as $x_M\to \infty$ with $x_Mp^- \simeq -l_2^-$}
\vspace{.2cm}

Finally, we consider the contraction of the $d$-quark fields in the MPD, which corresponds to $\pi^+ N \to \bar d + \gamma_L^* + X$.
Similarly to \eq{contr1} we have
\beq\label{contr2}
\bra{0}\bar\psi_d^\alpha(y_3)\psi_d^\beta(0)\ket{0} =
\left.\int\frac{dl^+\,d^2\bs{l}_\perp}{(2\pi)^3 2l^+}\theta(l^+)\,e^{-il\cdot y_3} \slashed{l}^{\beta\alpha}\,\right|_{l^-=l_\perp^2/l^+}
\eeq
After the integrals over $d^2\bs{l}_\perp$ and $dy_3^+$ and defining $y^- = y_1^- -y_2^- -y_3^-$ we get (with a minus sign from reordering the fermion fields)
\beqa\label{fdu3}
f_{d\bar u/p}&=&-\frac{1}{4(4\pi)^3}
\int  dy_1^- dy_2^- dy^- \int\frac{dl^+}{2\pi}\, l{^+}\theta(l^+)\,
 \bra{N(p)}\bar\psi_u(y^-)\gamma^+\,\psi_u(0) \ket{N(p)}\\
&\times& \exp\left[-\frac{i}{4}(y_1^- -y_2^-)(l_1^++{l_1'}^+)-\frac{i}{4}(y_1^- +y_2^-)(l_1^+-{l_1'}^+) -\frac{i}{2}(y_1^- -y_2^--y^-)(q^++l^+)\right] \nn
\eeqa
The integrals over $y_1^-,y_2^-$ set $l_1^+ = {l_1'}^+$ and $l^+ = -l_1^+-q^+ = -l_2^+$, giving
\beq
f_{d\bar u/p} = \delta(l_1^+
-{l_1'}^+)\, \frac{-l_2^+}{4\pi}\,\theta(-l_2^+) f_{\bar u(p}(-l_1^+/p^+)
\eeq
where the antiquark PDF is
\beq
f_{\bar u/
p}(-l_1^+/p^+) = -\frac{1}{8\pi}\int dy^- \exp(-\halft i y^- l_1^+)
\bra{N(p)}\bar\psi_u(y^-)\gamma^+\,\psi_u(0) \ket{N(p)}
\eeq
Now the dominant contribution of $C(x_B,x)$ in the cross section comes from its pole at $l_2^+ =0$, giving $\sigma \propto e_u^2$ as expected.

\end{document}